\documentclass[prl,twocolumn,aps,letterpaper, preprintnumbers,superscriptaddress]{revtex4}
\usepackage{textcomp}
\usepackage{hyperref}
\usepackage{verbatim}
\usepackage{amsmath}
\usepackage{latexsym}
\usepackage{revsymb}
\usepackage{yfonts}
\usepackage{ifthen}
\usepackage{tcolorbox}
\usepackage{natbib}
\usepackage{amsfonts}
\usepackage{amsmath}
\usepackage{amssymb}
\usepackage{amsthm}
\usepackage{graphicx}
\usepackage{bm}
\usepackage{bbm}
\usepackage{epsfig,color,amssymb}
\usepackage{subfigure}
\usepackage{amsfonts}
\usepackage{amscd}
\usepackage{amsmath}
\usepackage{multirow}
\usepackage{chemarrow}
\usepackage{dcolumn}
\usepackage{bm}
\usepackage{graphicx}
\usepackage{enumerate}
\usepackage{epsfig}
\usepackage{subfigure}
\usepackage{xcolor}
\usepackage{multirow}
\usepackage{ulem}
\usepackage{braket}
\usepackage{comment}
\usepackage{enumitem}
\usepackage{amsthm}
\usepackage{diagbox}
\usepackage{soul}
\renewcommand{\emph}[1]{\textit{#1}}

\newcommand{\YS}[1]{\textcolor{blue}{#1}}

\begin{document}
\title{Photonic Landau levels in a high-dimensional frequency-degenerate cavity}


\author{Jing Pan}
\affiliation{Key Laboratory of Photonic Control Technology (Tsinghua University), Ministry of Education, Beijing 100084, China }
\affiliation{State Key Laboratory of Precision Measurement Technology and Instruments, Department of Precision Instrument, Tsinghua University, Beijing 100084, China}
\author{Zhaoyang Wang}
\affiliation{Key Laboratory of Photonic Control Technology (Tsinghua University), Ministry of Education, Beijing 100084, China }
\affiliation{State Key Laboratory of Precision Measurement Technology and Instruments, Department of Precision Instrument, Tsinghua University, Beijing 100084, China}
\author{Yuan Meng}
\affiliation{Mechanical Engineering $\&$ Materials Science, Washington University in St. Louis, Saint Louis, Missouri 63130, United States}
\author{Xing Fu}\email{fuxing@mail.tsinghua.edu.cn}
\affiliation{Key Laboratory of Photonic Control Technology (Tsinghua University), Ministry of Education, Beijing 100084, China }
\affiliation{State Key Laboratory of Precision Measurement Technology and Instruments, Department of Precision Instrument, Tsinghua University, Beijing 100084, China}
\author{Yijie Shen}\email{yijie.shen@ntu.edu.sg}
\affiliation{Centre for Disruptive Photonic Technologies, School of Physical and Mathematical Sciences \& The Photonics Institute, Nanyang Technological University, Singapore 637371, Singapore}
\affiliation{School of Electrical and Electronic Engineering, Nanyang Technological University, Singapore 639798, Singapore}
\author{Qiang Liu}\email{qiangliu@mail.tsinghua.edu.cn}
\affiliation{Key Laboratory of Photonic Control Technology (Tsinghua University), Ministry of Education, Beijing 100084, China }
\affiliation{State Key Laboratory of Precision Measurement Technology and Instruments, Department of Precision Instrument, Tsinghua University, Beijing 100084, China}


\begin{abstract}
\noindent \textbf{Topological orders emerge in both microscopic quantum dynamics and macroscopic materials as a fundamental principle to characterize intricate properties in nature with vital significance, for instance, the Landau levels of electron systems in magnetic field.  Whilst, recent advances of synthetic photonic systems enable generalized concepts of Landau levels across fermionic and bosonic systems, extending the modern physical frontier. However, the controls of Landau levels of photons were only confined in complex artificial metamaterials or multifolded cavities. Here, we exploit advanced structured light laser technology and propose the theory of high-dimensional frequency-degeneracy, which enables photonic Landau level control in a linear open laser cavity with simple displacement tuning of intracavity elements. This work not only create novel structured light with new topological effects but also provides broad prospects for Bose-analogue quantum Hall effects and topological physics.}
\end{abstract}

\maketitle

\noindent Two-dimensional (2D) electron systems subjected to magnetic fields form Landau levels with quantized topological orders, which is regarded as a basic model in modern quantum physics and hatched several scientific branches such as topological insulators~\cite{moore2010birth,hasan2010colloquium,tokura2019magnetic,rhim2020quantum}, anyon collisions~\cite{bartolomei2020fractional,carrega2021anyons} and twistronics~\cite{uri2020mapping,bultinck2020mechanism}. Whilst, recent advances of synthetic quantum materials highlighted sophisticated photonic systems to emulate the properties of electrons~\cite{ozawa2019topological}, such as dielectric photonic crystals~\cite{fang2012realizing,khanikaev2013photonic,lustig2019photonic}, continuum photon fluids~\cite{carusotto2013quantum,situ2020dynamics}, and optical topology~\cite{gornik2021landau,harari2018topological,bandres2018topological}, merging the fermionic and bosonic systems and promoting physical frontier of quantum simulation~\cite{cirac2012goals}. Now, Landau levels can be further generalized beyond original fermionic systems into bosonic even anyonic systems~\cite{rechtsman2013strain,jamadi2020direct,devarakonda2021signatures,schine2016synthetic}, relying on elaborate photonic crystal designs~\cite{rechtsman2013strain,jamadi2020direct} or expensive superconductors~\cite{devarakonda2021signatures}. In addition to the solid-state systems, the Landau levels of free-space photons were also theoretically proposed and experimentally demonstrated~\cite{schine2016synthetic}, showing emergent breakthroughs in extending frontier of fundamental physics. For instance, the photonic Landau levels were utilized in topological characterization of electromagnetic and gravitational responses~\cite{schine2019electromagnetic} and photonic crystals~\cite{Barczyk2024}, exploration of light-matter interaction regarding exotic polaritons~\cite{jia2018strongly,clark2019interacting}, and emulation of quantum Laughlin matter~\cite{clark2020observation,corman2020light}.
However, the creation and control of photonic Landau level control is still a huge challenge. 
The only existing way to do this is using a spatially multifolded laser resonators, with require very precise and complicated control of non-planar geometry to manipulate cyclotron behavior of confined photons~\cite{schine2016synthetic}. The new physics and compact source for photonic Landau levels are highly desired for unlocking more potential practical applications, while the advanced structured light laser could provide a method to engineer twisted light and break the optical chiral symmetry~\cite{Forbes2024}.

Here, we propose a new mechanism to distribute photons (the structured modes) in Landau levels via a compact laser cavity, where the distribution of high-dimensional frequency-degenerate structured modes is created as the analogy of Landau levels. The crucial step of our method is designing the effective magnetic fields for the structured modes by exploiting intracavity astigmatic mode convertor to twist structured light modes in the cavity. Via simply manipulating the linear displacement of cavity elements, the structured laser modes can be emitted in discrete frequency degenerate states, which fulfills the same Hamiltonian in specific conditions of Landau energy levels. We also experimentally observe the Landau quantization corresponding to a series of exotic orbital angular momentum (OAM) laser modes in different photonic Landau energy levels. Our method could distribute structured modes with different orders in photonic Landau levels flexibly by only tuning the linear displacement of cavity elements. Our work unveils the possibility to map various nontrivial topological effects into the structured modes, such as Aharonov-Bohm effect, topological edge states, Zeeman effect and quantum Hall effect, providing a physical insights for electron-photon analogies.

\begin{figure*}[t!]
	\centering
	\includegraphics[width=0.8\linewidth]{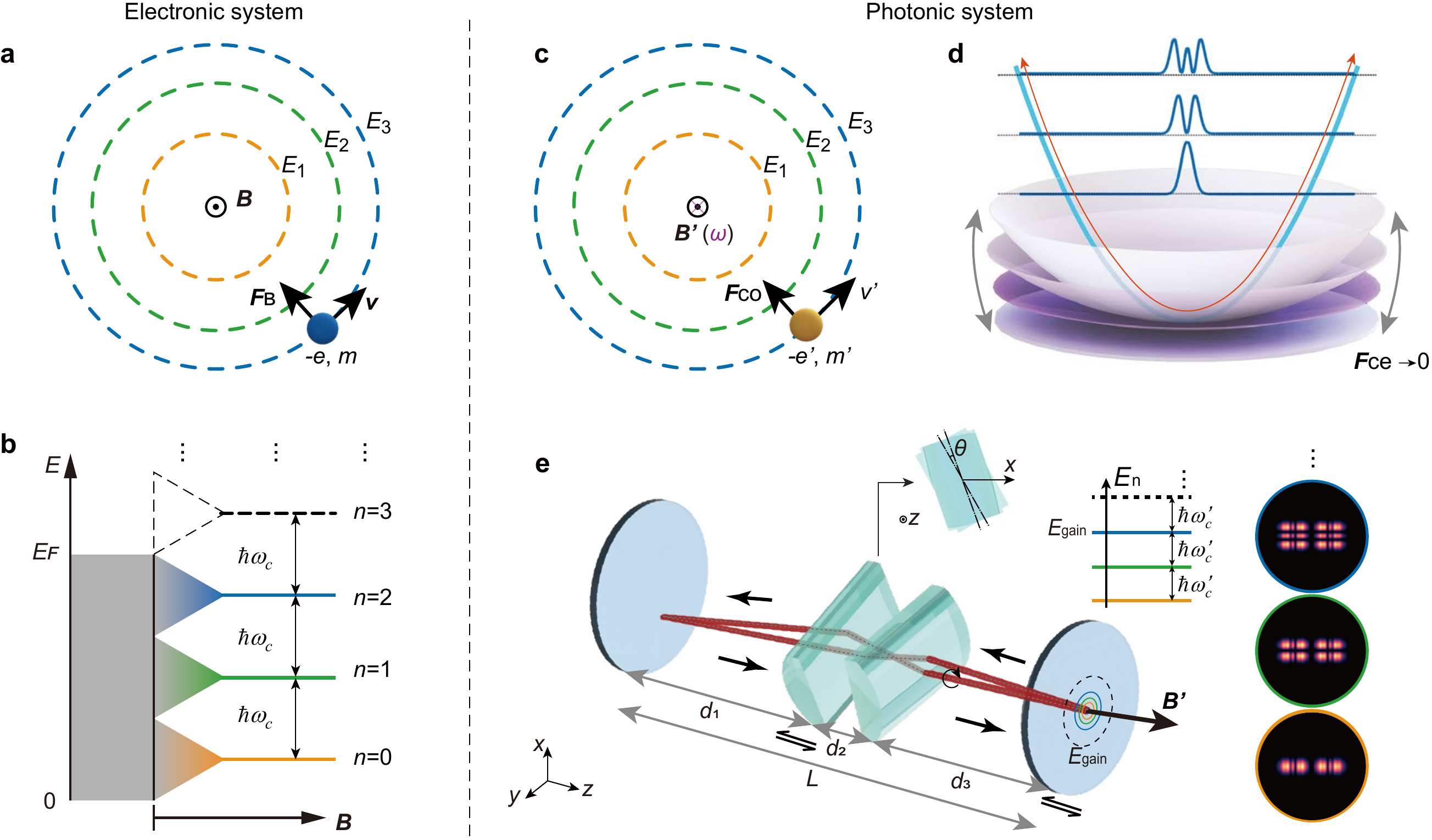}
\caption{\textbf{Laudau levels of electrons and photons.} \textbf{a}, In a 2D electronic system under a uniform magnetic field $\bm{B}$ (perpendicular to the paper pointing outwards), an electron with charge of $-e$ and mass of $m$ can rotate along discrete circle orbits (orange, green, and blue dashed lines, with energy of $E_{1}$, $E_{2}$ and $E_{3}$, respectively) at velocity of $\bm{v}$ driven by Lorentz force of $\bm{F}_{\bm{B}}$. \textbf{b}, With the increasing of magnetic field, discrete energy levels $E_{1}$, $E_{2}$ and $E_{3}$ form with constant spacing of $\hbar\omega_{c}$ and quantum numbers, $n=0,1,2,3,\cdots$, where the highest number is subjected to the Fermi energy $E_F$. \textbf{c}, In the uncharged particle system, the reference system is a rotating system with the angular velocity $\omega$ with direction perpendicular to the paper pointing inwards, in which, a particle rotates at velocity of $\bm{v}'$ along the circle orbits is driven by Coriolis force of $\bm{F}_{\text{co}}$, analogue to the Lorentz force in \textbf{a}. According to the similarity to the quantized energy levels of electron cyclotron orbits in \textbf{a}, the effective charge $-e'$ and mass $m'$ of the uncharged particle, and the effective magnetic field $\bm{B}'$ (perpendicular to the paper pointing outwards) can be derived. \textbf{d}, Besides the Coriolis force, the centrifugal force $\bm{F}_{\text{ce}}$ is also introduced in the rotating reference system. As the centrifugal force is gradually converging to $0$, the energy levels evolve from the independent oscillators to an energy-degenerate state. \textbf{e}, As the photonic system, the rotating system is introduced by the designed laser cavity with two nonparallel cylindrical lenses inside (angle between two lines generatrices are marked as $\theta$). By tuning the displacements of elements (indicated by ``$\rightleftharpoons$''), the effective centrifugal force $\textit{\textbf{F}}_{\text{ce}}$ can be tuned to $0$ and Landau levels can be excited. The wave functions of these Landau levels $\textit{{E}}_{n}$ corresponds to various laser mode patterns, shown in the right orange, green, blue circles, and the highest energy level is subjected to the gain aperture $\textit{{E}}_{\text{gain}}$, analogue to the Fermi Energy $E_F$ in \textbf{b}.} 
	\label{f1}
\end{figure*}

\section{Results}
\noindent
\textbf{Concept.} When applying a uniform magnetic field to an electron system, the electrons can only occupy discrete cyclotron orbits, i.e. Landau energy levels, and this process is called Landau quantification. For instance, an electron rotating with the velocity {$\bm{v}$} in the magnetic field $\bm{B}$ is driven by the Lorentz force pointing to the center of the orbits, see Fig.~\ref{f1}\textbf{a}. The Hamiltonian of this system is given by $H = {\left( {\bm{p} - e\bm{A}}\right)^{2}}/{(2m)}$ based on the Fock–Darwin model \YS{\cite{fock1928bemerkung,darwin1931diamagnetism}}, where $\bm{p}$ is the canonical momentum operator, $\bm{A}$ is the electromagnetic vector potential, $-e$ and $m$ are the charge and mass of the electrons, respectively. With the different values of cyclotron angular velocity $\omega_{c}$, the electrons rotate on discrete orbits corresponding to different energy levels $E_{n}$ where $E_{n}=(n+1/2)\hbar\omega_{c}$ ($n=0,1,2,\cdots$). The spacing between any pair of adjacent energy levels is a constant $\hbar\omega_{c}$ where $\omega_{c} = eB/m$, as shown in Fig.~\ref{f1}\textbf{b}, and the highest energy level is determined by Fermi energy, $E_F = \hbar^2 k_F^2 / (2m)$ ($k_F$ is the Fermi wave vector), of the electron system. 

The fundamental model can be generalized from electron systems to uncharged particle systems~\cite{schine2016synthetic}. For a charge-neutral particle with an introduced rotating coordinate system in cyclone velocity $\omega$ (Fig.~\ref{f1}\textbf{c}), the particles with relative rotational motion of this coordinate system are driven by the Coriolis force $\bm{F}_{co}$, which can be analogous to the equivalent Lorentz force with a synthetic magnetic field $\bm{B'}$. Thus, the synthetic Landau levels can be achieved. Besides the Coriolis force, the centrifugal force $\bm{F}_{ce}$ is also introduced with the rotating coordinate system. The Hamiltonian can thus be written as $H = {\left( {\bm{p} - e'\bm{A'}} \right)^{2}}/{(2m)} - \bm{\Omega} \cdot \bm{L}$, where $\bm{A'}$ is the synthetic magnetic field potential corresponding to $\bm{B'}$, $\bm{\Omega}$ is the angular velocity, $\bm{L}$ is the angular momentum, and  $-\bm{\Omega} \cdot \bm{L} = \frac{1}{2}m\omega_{\text{trap}}^{2}\bm{r}^{2}$ corresponds to the centrifugal force, where $\omega_{\text{trap}}$ is the trapping frequency (corresponding to the marked $\omega$ physically in Fig.~\ref{f1}\textbf{c}) related with the centrifugal force, and $r$ is the particle’s transverse position vector. The introduction of the centrifugal force changes the energy levels of the system and keeps them away from degeneracy. When the $\bm{F}_{ce}$ and $\omega_{\text{trap}}$ are gradually converging to 0, the Hamiltonian becomes the mathematically same with that of the electronic system, and the harmonic potential is flattened in Fig.~\ref{f1}\textbf{d} to achieve the Landau levels. Noted that the $\omega_\text{trap}$ is arsed by the centrifugal force, which acts against the hypothetical centripetal force. Therefore, $\omega_\text{trap}$ could converge to $0$ if a real force is introduced to act as the centripetal force.

The photonic system as a charge-neutral particle system, its synthetic Landau level can be achieved in a judiciously designed laser cavity as Fig.~\ref{f1}\textbf{e}. In the cavity, a pair of cylindrical lenses (with an angle difference $\theta$ between their principal axes) was introduced, which acted as an intracavity astigmatic convertor to excite the high-order structured modes~\cite{pan2020index,pan2023multiaxial}. The excited modes are ray-wave geometric beams that the intensity distribution is located on several discrete rays, which could be represented by the schematic arrows, as shown in Fig.~\ref{f1}\textbf{e}. The oscillating laser rays would rotate around the principal axis of the cavity, then a rotating reference system could be introduced to analyze the oscillating rays in the cavity. The hypothetical centripetal ${\bm{F}}_\text{ce}$ could be introduced in the rotating reference system. The hypothetical centripetal force could be regarded as the effective Lorentz force in an electron system, where the effective magnetic field $\bm{B'}$ is along the principal axis of the cavity. The cavity with the intracavity astigmatic convertor could be regarded as a harmonic oscillator, where the eigenmodes corresponding to various energy levels are shown in Fig.~\ref{f1}\textbf{d}. The hypothetical centrifugal force ${\bm{F}}_\text{ce}$ can be tuned to $0$ by manipulating the cavity mirrors and the cylindrical lenses, as shown in the black double arrows $\rightleftharpoons$ in Fig.~\ref{f1}\textbf{b}, to achieve the synthetic Landau level. The excited modes could be decomposed to a superposition of frequency-degenerate eigenmodes, i.e. several eigenmodes with the same frequency but different indices. The excited modes are the analogy of the quantum states in the synthetic Landau levels. The quantum wave functions in different synthetic energy levels $\textit{{E}}_{n}$ correspond to the excited mode in different frequency-degenerate states, which could be tuned by manipulating the off-axis displacement of cavity elements. The tunable excited modes are shown in the right orange, green, blue circles in Fig.~\ref{f1}\textbf{e}, where the gain aperture limits the highest energy level occupied by photons $\textit{{E}}_\text{gain}$, as the effective Fermi energy.

\emph{Electron-photon analogy}. For a clear view, the electron-photon analogies between the quantized electrons and the structured modes are summarized in the Table.~\ref{tab1}. The motion of electrons in the magnetic field $\bm{B}$ is driven by the Lorentz force, where the cyclotron orbits are quantized corresponding discrete energy levels, i.e. the Landau level $E_{n}$. The difference of Landau level is a constant related to the frequency space $\omega_{c}$. While the motion of the photons (the structured modes) in the synthetic magnetic field $\bm{B}$ (the high-dimensional frequency-degenerate cavity) is driven by the hypothetical centripetal force, where the frequency of the excited modes are quantized corresponding discrete energy levels $E_{n}$, i.e. the analogy of the Landau level. The difference of photonic energy level is a constant related to the frequency space $\omega_{c}'$. The core point of this analogy is the discrete energy level, as discussed in follows.

\begin{table}[htbp]
\renewcommand\arraystretch{1.5}
\centering
\caption{The electron-photon analogies between the quantized electrons and the structured modes.}
\begin{tabular}{|c|c|c|c|}
\hline
\multicolumn{1}{|c|}{ Term } & \multicolumn{1}{|c|}{ Quantized electrons } & \multicolumn{1}{|c|}{ Structured modes } \\
\hline
Particle & Electrons & Photons \\
Force & The Lorentz force & The centripetal force\\
Magnetic field & Real field $\bm{B}$ & Synthetic field $\bm{B'}$ \\
Frequency & $\omega_{c} = eB/m$ & $\omega_{c}'$ \\
Energy level & $E_{n}=(n+1/2)\hbar\omega_{c}$ & $E_{n}=(n+1/2)\hbar\omega_{c}'$ \\
\hline
\end{tabular}
\label{tab1}
\end{table}

\begin{figure*}[t!]
	\centering
	\includegraphics[width=\linewidth]{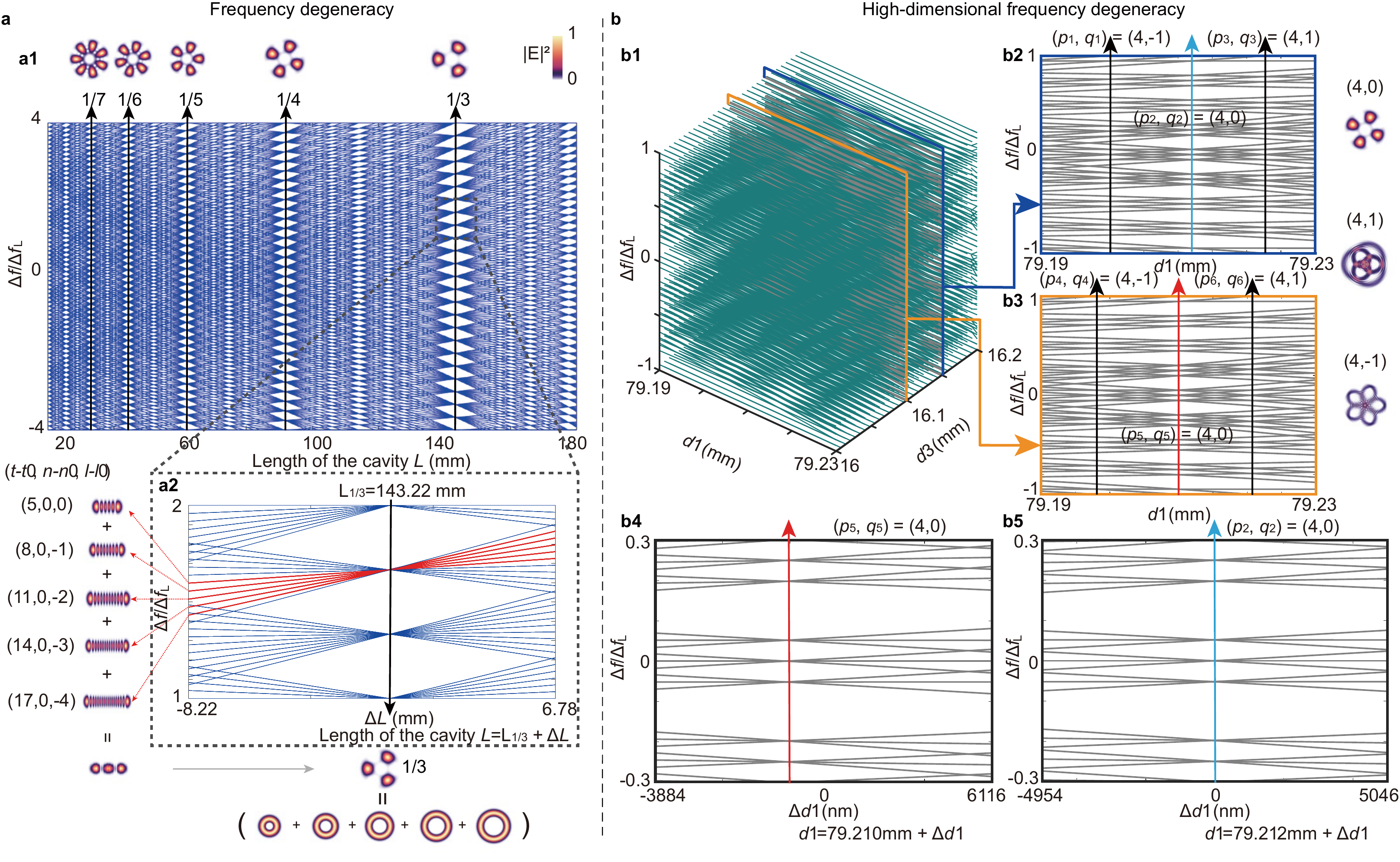}
	\caption{\textbf{Frequency degeneracy.} \textbf{a1}, 1D frequency spectrum versus cavity length ($L$) of a conventional confocal cavity, where a series of frequency-degenerate states can be observed with theoretical emitting OAM geometric modes marked correspondingly, by tuning $L$. \textbf{a2}, The zoom-in of a special degenerate state of $P/Q=1/3$, where the frequency lines correspond to various eigenmodes with different transverse and longitudinal mode indices (selective HG mode patterns corresponding to the highlighted red lines are marked). At a frequency-degenerate state (intersection point of a set of eigenmodes), the set of eigenmodes can be superposed together into a coherent state (left-bottom inset). By introducing OAM to the modes (converting HG bases into LG), the OAM geometric modes can be generated with exotic multi-petal patterns, and such patterns can be used to identify the types of degenerate states (different $P/Q$ ratio). \textbf{b}, High-dimensional frequency spectrum \textbf{b1}, versus length parameters $d_1$ and $d_3$, which are tuned respectively, to generate different frequency-degenerate states. The frequency lines correspond to the eigenmodes with different 2D transverse mode and a longitudinal mode indices, and cross sections in the blue and orange frames are selected. \textbf{b2}, The cross section at $\textit{d}3=16.15$~mm shows different frequency-degenerate states with $(p_{1},q_{1})=(4,-1)$, $(p_{2},q_{2})=(4,0)$ and $(p_{3},q_{3})=(4,1)$ at specific values of $d_1$ marked by the arrows. \textbf{b3}, The cross section at $\textit{d}3=16.1$~mm where the different frequency-degenerate states with $(p_{4},q_{4})=(4,-1)$, $(p_{5},q_{5})=(4,0)$ and $(p_{6},q_{6})=(4,1)$ at specific values of $d_1$ marked by the arrows. The LG-based superposed geometric modes corresponding to different degenerate families were marked on the right. \textbf{b4,b5}, The zoom-in inserts of figures \textbf{b2} and \textbf{b3} at degenerate states, $(p_{5},q_{5})=(4,0)$ and $(p_{2},q_{2})=(4,0)$, marked by the red and blue arrows.} 
	\label{f2}
\end{figure*}
\begin{figure*}[t!]
	\centering
	\includegraphics[width=0.825\linewidth]{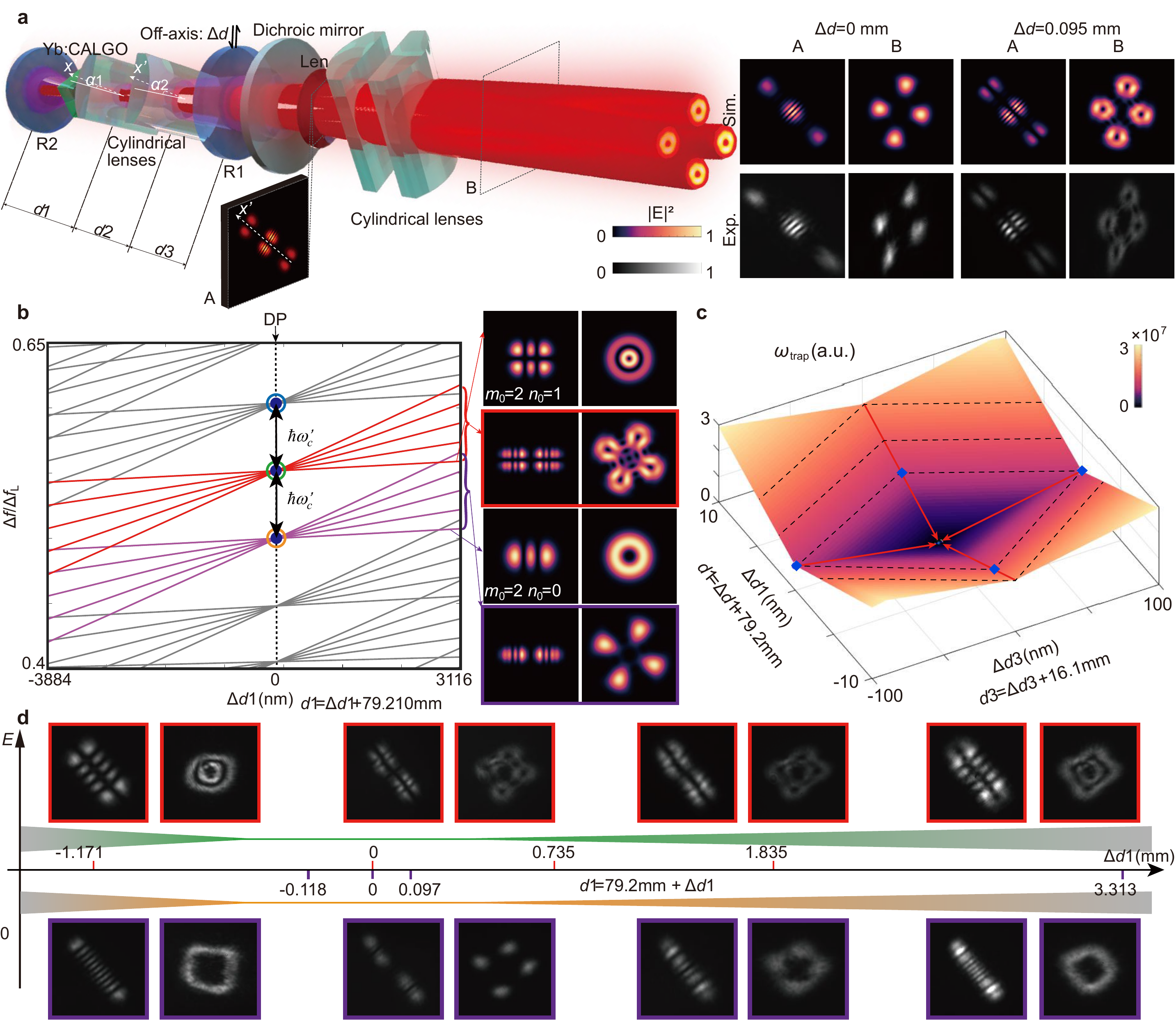}
	\caption{\textbf{Experimental results.} \textbf{a}, Our experimental cavity design for synthetic Landau levels, which is built with two cavity mirrors (R1 and R2, Radius of curvature are $R1$ and $R2$), the gain crystal Yb:CALGO, and a pair of cylindrical lenses (focal lengths: $f$) with rotating angles $\alpha_1$ and $\alpha_2$ (the $x'$-direction parallel to index $t$ of generated HG modes set as the 0 degree baseline. The distances among R2, the two cylindrical lenses, and R1 are $d_1$, $d_1$, and $d_3$, respectively. The synthetic magnetic field can be controlled by tuning $d_1$ and $d_3$. The orders of HG modes can be controlled by tuning $\omega_{\text{trap}}$. Another mode converter composed of a pair of cylindrical lenses outside the cavity is used to convert the OAM of output mode, from A to B, with simulated and experimental patterns shown in the right insets. \textbf{b}, A cross-section of high-dimensional frequency spectrum of the cavity is at $d_3=16.1$~mm, and at the frequency-degenerate position (DP), different groups of lines (red lines and purple lines) correspond to different Landau levels which can be tunable with $\omega_{\text{trap}}$, and their related modes with different indices $n_{0}$ shown on the right (including a single eigenmode with one line and a superposed mode with a set of lines). \textbf{c}, $\omega_{\text{trap}}$ evolves with the tunability of $d$3 and $d$1, which represents the effect of Centrifugal force. The zero point of $\omega_{\text{trap}}$ is the Landau level point, also the frequency-degenerate point (DP). \textbf{d}, Experimental Landau level control: the output frequency-degenerate modes of DP at $\Delta{d}=0$ corresponding to patterns in the $3^\text{rd}$ (HG-based) and $4^\text{th}$ (LG-based) columns, around which the superposed modes gradually evolve out of degeneracy by tuning $d_{1}$ as the $1^\text{st}$, $5^\text{th}$, $7^\text{th}$ columns (HG-based) [the $2^\text{nd}$, $6^\text{th}$, $8^\text{th}$ columns (LG-based)], corresponding to the $1^\text{st}$, $3^\text{rd}$ and $4^\text{th}$ red (purple) vertical lines on the horizontal axis. The patterns of the lower (upper) line in the purple (red) frames represent the modes around $E_{n=0}$ ($E_{n=1}$) with index $n_{0}=0$ ($n_{0}=1$) corresponding to the purple (red) vertical lines on the horizontal axis, achieved by the different off-axis of R1.} 
	\label{f3}
\end{figure*}

\noindent
\textbf{High-dimensional frequency-degenerate cavity.} The photonic Landau levels could be achieved in experiment by controlling the designed frequency-degeneracy condition of the structured laser cavity. The frequency spectrum of the excited structured modes in the cavity is expressed as $f_{n,m,l} = f_z [l +(n+1/2)f_x/f_z+(m+1/2)f_y/f_z]$, where $(n,m,l)$ and $(f_x,f_y,f_z)$ are the mode indices and the frequency spaces along $x, y, z-$ directions.
The frequency-degeneracy condition refers that the modes with different indices have the same frequency, i.e. $\Delta f_{n,m,l} = f_{n,m,l} - f_{n_0,m_0,l_0} = 0$, where $(n_0,m_0,l_0)$ are the initial mode indices. As a simple case that $f_x = f_y = f_0$, Fig.~\ref{f2}\textbf{a} shows the frequency spectrum corresponding to a plano-concave cavity~\cite{chen2013exploring,tung2016fractal,shen2020structured}, where the laser mode frequency spacing is given as $\Delta f_{n,m,l} = [(n-n_0)+(m-m_0)]f_0 + (l-l_0)f_z $. The frequency-degeneracy condition requires $f_0/f_z = P/Q$ ($P$ and $Q$ are coprime) and $l_0 - l = PK$, $n-n_0 = pK$, $m-m_0 = qK$ and $p+q= Q$, where $K$ is an integer and $P/Q$ is determined by the cavity length $L$ as $L=R\tan(\pi P/Q) /2$, $R$ is the curvature of the concave mirror. $\frac{\Delta f_{0}}{\Delta f_{z}}$ can also be obtained based on the ABCD matrix (Supplementary Material).

Each line in Fig.~\ref{f2}\textbf{a1} represents the frequency evolution of a certain eigenmode by tuning the cavity length. At the special cavity lengths, marked by the black arrows in Fig.~\ref{f2}\textbf{a1}, several lines converge at one point, which means the frequency-degeneracy has occurred. These eigenmodes corresponding to the converging lines have the same frequency, could be coherently superposed. The superposed modes usually exhibit that the intensity distribution are located on several discrete rays, which are usually called ray-wave beams~\cite{shen2021rays,chen2021laser}. For instance, the simulated geometric modes (superposed by Laguerre-Gaussian (LG) eigenmode) are inserted in Fig.~\ref{f2}\textbf{a1}. The zoom-in of the frequency spectrum at a special cavity length with $P/Q=1/3$ is shown in Fig.~\ref{f2}\textbf{a2}, where red lines are selected as the example. These discrete lines corresponds to a collection of eigenmodes with various indices but the same frequency. The indices spaces $(n-n_0,m-m_0,l_0-l)$ are also labeled in Fig.~\ref{f2}\textbf{a2}. The superposed eigenmodes based on HG and LG modes are exhibited in the left and right areas, which can be converted to each other via an astigmatic mode converter.

\emph{The photonic Landau levels.} For the photons in a frequency degenerate cavity with $f_x = f_y = f_0$, their energy level could be expressed as $E_{N,l_0} = [l+(N+1)P/Q]]f_z$ where $N = n_0 + m_0$. The energy level difference between two adjacent frequency degenerate modes is $\Delta E = f_z$ for a fixed $N$ or $\Delta E = (P/Q)f_z$ for a fixed $l_0$, i.e. the energy level of the frequency-degenerate is discrete with a constant difference, analogous to the Landau levels, thus called the photonic Landau levels.

Besides, the photonic Landau levels can be generalized into high dimensions in our newly designed cavity with the intracavity mode convertor, where the displacement of various cavity elements can provide multiple synthetic degrees of freedom (DoFs). Figure~\ref{f2}\textbf{b} shows the example that the frequency degenerate spectrum versus two synthetic DoFs, where $d_1$ and $d_3$ are the tunable lengths labeled in Fig.~\ref{f1}\textbf{e}, respectively. In high-dimensional degeneracy, $f_x \neq f_y$. Thus, the eigenmode frequency spaces is $\Delta f_{n,m,l} = (n-n_0)f_x+(m-m_0)f_y + (l-l_0)f_z $ where $f_x = f_0 + q\Delta f$ and $f_y = f_0 - p\Delta f$. ${\Delta f_{x}}/{\Delta f_{z}}$ and ${\Delta f_{y}}/{\Delta f_{z}}$ can be calculated based on the generalized ABCD matrix theory (Supplementary material), corresponding to the tunable $d_{1}$ and $d_{3}$. To illustrate the relationship between the frequency spaces and ($d_{1}$, $d_{3}$), the high-dimensional frequency spectrum is shown in Fig.~\ref{f2} b1, where $d_{1}$ ranges from 79.19~mm to 79.23~mm and $d_{3}$ ranges from 16~mm to 16.2~mm in our experimental setup (Supplementary material). To exhibit the frequency lines clearly, the subspace degenerate spectra at $d_{1}=16.10$~mm and $d_{1}=16.15$~mm are shown in Figs.~\ref{f2} b2 and b3, corresponding to the orange and blue frames in Figs.~\ref{f2} b1. The superposed structured modes at the $d_{3}=16.15$~mm and $d_{3}=16.1$~mm with some certain $d_{1}$ (marked by arrows in Fig.~\ref{f2} \textbf{b2} and \textbf{b3}) are decomposed into several eigenmodes under the frequency-degeneracy of $(p,q)=(4,0)$, $(4,-1)$, and $(4,1)$, respectively. The simulated results of the superposed structured modes based on LG modes are shown in the right column as examples. The zoom-in figures around the frequency degeneracy with $(p,q)=(4,0)$ (the red and blue arrows) are shown in Figs.~\ref{f2} \textbf{b4} and \textbf{b5}, respectively.

\emph{The high-dimensional photonic Landau levels.} For the generalized case that $f_x \neq f_y$, their energy level could be expressed as $E_{n_0, m_0,l_0} = (n_0+\frac{1}{2})f_x + (m_0+\frac{1}{2})f_y + l_0 f_z$. The energy level difference between two adjacent frequency degenerate modes is $\Delta E = f_x, f_y, f_z$ for a fixed $(m_0, l_0)$, $(n_0, l_0)$, and $(n_0, m_0)$, respectively, akin to the Landau levels.
\\[4pt]

\noindent
\textbf{Experiment results.} We assembled the cavity shown in Fig.~\ref{f1} e where the two intracavity cylindrical lenses are set in the angle of $\alpha_{1}$ and $\alpha_{2}$, respectively. The distances $d_{1}$ (between the first cylindrical lens and R2) and $d_{3}$ (between the second cylindrical lens and R1) are tunable to achieve the frequency-degenerate superposed modes, as shown in Fig.~\ref{f3} a. The output modes from the cavity are frequency-degenerate superposed HG modes (at the position A in  Fig.~\ref{f3} a), which would be transformed into frequency-degenerate superposed LG modes (at the position B in  Fig.~\ref{f3} b) by the mode conversion (the extracavity cylindrical lenses). The off-axis displacement of the first intracavity cylindrical lens would tune the index range of superposed HG modes in $x'$-direction, while the off-axis displacement of R2 would tune the index range of superposed HG modes in $y'$-direction. The frequency-degenerate superposed HG (LG) modes with the tunable $y'$-direction index under different $\delta d$ are shown in the right column of Fig.~\ref{f3} a, where the top and bottom raw correspond to the simulation and experimental results, respectively.

The frequency spectrum with the parameters of this experimental cavity in  Fig.~\ref{f3} b shows the lines corresponding to the HG eigenmodes with different $x'$-direction indices intersect but those corresponding to the HG eigenmodes with different $y'$-direction don't intersect, which means these sets of HG eigenmodes are frequency-degenerate with $x'$-direction indices, but not with $y'$-direction indices. Different sets of HG eigenmodes superpose different HG (LG)-based frequency-degenerate modes as the right inset of Fig.~\ref{f3} b, which represent different Landau levels, and the Landau level can be tunable with the off-axis of $\delta d$ in experiment. Moreover, with $d_{3}=16.1$~mm and $d_{1}=79.2$~mm, the $\omega_{\text{trap}}$ representing the trapping frequency related with the centrifugal force is {zero} (Supplementary material), as shown in Fig.~\ref{f3} c. Both Fig.~\ref{f3} b and c show the modes would be out of degeneracy when the cavity parameters deviate from frequency-degenerate position (DP), and the experimental results demonstrate this as Fig.~\ref{f3}d. When $d_{1}$ is tunable away from the DP, the modes evolve into the quasi-degenerate modes as HG (LG)-based modes in the third two columns and the forth two columns, as well as HG (LG) eigenmodes in the first two columns. The lower and upper rows of experimental modes represent the lower and higher Landau level states, as discussed in Fig.~\ref{f3}a and b.
\\[4pt]

\section{Discussion}
Landau levels lay the most fundamental topological orders for modern electronic and solid-state physics, especially, with profound implications for topological insulators, photonic crystals, and optoelectronic materials. While, the emerging research area of manipulating photonic Landau levels is still in its infancy, which is a concept that extends the Landau levels from electronic systems to photonic systems, towards the unification of fermionic, bosonic, and anyonic physics. Our work proposes a highly flexible and simple setup to create and manipulate photonic Landau levels of photons, unlocking their practical applications. Therefore, our work not only produces new kinds of structure light fields with unusual topological features, but also opens up new possibilities for studying quantum Hall effects, topological phase, and quantum simulation in a new platform of bosonic systems.

Our linear degenerate cavity method to control photonic Landau levels, in contrast to the prior work, is more stable and flexibly tunable.  Due to the extended degrees of freedom, we can precisely control photonic Landau levels and solve the trade-off between trap stability and geometry modification. However, because of the spatial limit of the linear cavity, it is challenging to simulate large-scale Landau levels in our current setup. Fortunately, recent advanced methods in higher-dimensional structured light manipulation promised many extended dimensions in order to access larger-range topological orders~\cite{forbes2021structured,he2022towards,shen2022topological}. For instance, the methods of intracavity spatial light modulator (SLM) and coupled cavity were used to simulate parity-time (PT) symmetry breaking~\cite{yu2021spontaneous,arwas2022anyonic}. We believe combining intracavity SLMs and degenerate laser could be an effective method to explore advanced solid-state physics and more properties of large-scale photonic crystals but in structured laser cavity, such as PT symmetry, non-Hermitian, and chiral edge states.



\section{Methods}
\noindent
\textbf{Frequency-degenerate mode laser source.} Our synthetic Landau level cavity is designed with two mirrors and two cylindrical lenses, among which, two cylindrical lenses are introduced to break the symmetry and achieve the output of 2D modes. To create a rotating system for the photons in the cavity, the generatrices of the pair of intracavity cylindrical lenses are not parallel. Specifically, the cavity is generated with cavity mirrors R1 (Radius of curvature: $R_{1}$) and R2 (Radius of curvature: $R_{2}$), the gain crystal Yb: CALGO, and a pair of cylindrical lenses (Both focal lengths: f) with rotating angulars $\alpha_{1}$ and $\alpha_{2}$. The distances between R2, the first cylindrical lens, the second cylindrical lens and R1 are $d_{1}$, $d_{2}$ and $d_{3}$, respectively. The synthetic magnetic field can be tuned with the distance $d_{1}$ between R2 and the first cylindrical lens and the distance $d_{3}$ between the second cylindrical lens and R1 by moving the first cylindrical lens and R1 along the axis, respectively. The high-order HG eigenmodes can be generated with the off-axis of the first cylindrical lens in cavity and R2.
\\[4pt]
\bibliographystyle{naturemag}

\noindent
\textbf{Acknowledgments} \\ This work is supported by Beijing Natural Science Foundation (JQ23021); Y. Shen acknowledges the support from Nanyang Technological University Start Up Grant, Singapore Ministry of Education (MOE) AcRF Tier 1 grant (RG157/23), MoE AcRF Tier 1 Thematic grant (RT11/23), and Imperial-Nanyang Technological University Collaboration Fund (INCF-2024-007).
\\[8pt]
\noindent
\textbf{Author contributions} \\ Y. S. conceived the idea and wrote the first version of manuscript. J. P. designed and performed the experiment. J. P. and Z. W. contributed to the theoretical modeling and analyzed the experimental data. X. F., Y. S., and Q. L. supervised the project. All authors took part in discussions, interpretations of the results, and revisions of the manuscript.
\\[8pt]
\noindent
\textbf{Competing interests}\\
The authors declare no competing interests.

\bibliography{sample}

\end{document}